\documentclass[amssymb,aps,floatfix,showkeys,showpacs]{revtex4}
\usepackage{graphicx}
\usepackage{amsmath}

\begin{document}

\title{Nonlinear Dependence of the Phase Screen Structure Function on the Atmospheric Layer Thickness}

\author{Richard J. Mathar} 
\email{mathar@strw.leidenuniv.nl}
\homepage{www.strw.leidenuniv.nl/~mathar}
\affiliation{Leiden Observatory, Leiden University, P.O. Box 9513, 2300 RA Leiden, The Netherlands}


\date{\today}

\begin{abstract}
The
phase structure function accumulated by two parallel rays
after transmission
through a layer of turbulent air is best known by a proportionality to the 5/3rd power of the lateral distance in the aperture,
derived from an isotropic Kolmogorov spectrum of the refractive index.
For a von-K\'arm\'an spectrum of the refractive index,
a dependence involving a modified Bessel function
of the ratio of the distance over the outer scale is also known.

A further standard proposition
is a proportionality to the path length through
the atmospheric layer. The manuscript
modifies
this factor through a refined calculation of an
integral representation
of the structure function.
The correction establishes a sub-linearity as the lateral distance grows in proportion to
the layer thickness;
it is more important for large than for small outer scales.

\end{abstract}

\maketitle

\section{Scope}

\subsection{Planar Phase Screen Terminology}

A phase screen characterizes the distribution of optical path
lengths in the aperture of a receiver
depending on the inhomogeneity of the refractive index along the path
through the atmosphere up to the emitter
\citep{HufnagelJOSA54}.
The phase 
$\varphi({\bf r})$ in the receiver's pupil plane is a function of
the two-dimensional spatial vector ${\bf r}$. The phase is the line integral
of the product of the vacuum wave number $\bar k=2\pi/\lambda$ by the index
of refraction $n$ along the path through the atmosphere at the wavelength
$\lambda$ of the electromagnetic spectrum.

The seeing quality, figures of merit like the Strehl ratio, and the
requirements on adaptive optics systems aiming to straighten these
wave fronts refer to these phase screens.  The current work is a contribution
to an accurate description of the phases, the measured quantities,
in terms of  parameters that describe the turbulence
(structure constant and outer scale length) and in terms of geometric parameters which 
determine the sampling (optical path length of the rays and distance between these).

In the remainder of this section, the manuscript recalls
how a model of isotropic turbulence is set up, then reviews
in Section
\ref{sec.2}
how this leads through a procedure of integration
of the refractive index along two parallel rays to a standard formula
of the phase structure function of von-K\'arm\'an turbulence. This is redone in Section
\ref{sec.3}
on a more accurate basis that keeps the layer thickness parameter consistent
throughout the computation, with the main result that a finite layer
thickness generates a
smaller phase structure function than with the
standard formula.

\subsection{Three-Dimensional Turbulence Spectrum}

This section summarizes the basic theory which links the structure
functions of three-dimensional indices of refractions to phase
covariances measured in pupil planes.

The structure function of refractive indices $n$ is defined as
the expectation value of squared differences if measured at two points
separated by a 3-dimensional vector $\Delta \mathbf{r}$,
\begin{equation}
{\cal D}_n \equiv \langle[n(\mathbf{r})-n(\mathbf r+\Delta\mathbf{r})]^2\rangle
.
\end{equation}
Assuming that the expectation value of the mean does not depend on $\mathbf r$,
binomial expansion rephrases this as a background value minus two times
a correlation function,
\begin{equation}
{\cal D}_n(\Delta \mathbf r) = 2 \langle n^2\rangle-2\langle n(\mathbf r)n(\mathbf r +\Delta\mathbf{r})\rangle
.
\label{eq.Dn}
\end{equation}
Its Fourier Transform as a function of spatial frequencies $f$ is
\begin{equation}
{\cal D}_n(\Delta \mathbf r) = 2 \int \Phi_n(f)[1-\cos(2\pi \mathbf f\cdot \Delta \mathbf r)]d^3f
\label{eq.DofPhi}
.
\end{equation}
The imaginary part $\propto i\sin(2\pi\mathbf f\cdot \Delta \mathbf r)$ vanishes as we
assume that the structure function has even parity: ${\cal D}_n(\Delta \mathbf r)={\cal D}_n(-\Delta \mathbf r)$.
Other notations emerge if the factor $2\pi$ is absorbed into the spatial
frequency or if the normalization of the Fourier Transform is chosen differently.

We focus on the von-K\'arm\'an model of an isotropic power spectrum with zero inner scale,
a prefactor $c_nC_n^2$ ---which
is split into a constant $c_n$ and a structure constant $C_n^2$---, an outer scale
wave number $f_0$ representing the inverse of the outer scale length, and a power
spectral index $\gamma$,
\begin{equation}
\Phi_n(f) =c_n C_n^2(f^2+f_0^2)^{-(\gamma+3)/2}.
\end{equation}
The Kolmogorov Model is obtained by
setting $\gamma=2/3$ and setting the outer scale to infinity, i.e., $f_0=0$.
This $\Phi$ is inserted in (\ref{eq.DofPhi}). Polar coordinates are introduced in
wave number space with Jacobian determinant $f^2\sin(\theta_f)$, and $\mathbf f\cdot \Delta \mathbf r=fr\cos \theta_f$.
Integration over the azimuth
in these coordinates yields a factor $2\pi$, and the integration
over the colatitude $\theta_f$ introduces a shape factor with a spherical
Bessel function $j_0(x)\equiv \sin(x)/x$,
\begin{eqnarray}
{\cal D}_n(\Delta \mathbf r)
&=& 4\pi c_nC_n^2 \int \frac{f^2}{(f^2+f_0^2)^{(\gamma+3)/2}}
\times
[1-\cos(2\pi f r\cos \theta_f)]df \sin(\theta_f) d\theta_f
\\
&=& 8\pi c_nC_n^2 \int_0^\infty \frac{f^2\,df}{(f^2+f_0^2)^{(\gamma+3)/2}}
[1-j_0(2\pi f r)].
\label{eq.D1d}
\end{eqnarray}
We require that ${\cal D}_n$ does not depend on $c_n$ in the Kolmogorov limit:
\begin{equation}
\lim_{f_0\to 0}{\cal D}_n \to C_n^2 r^\gamma,
\label{eq.Dnf0}
\end{equation}
and
this fixes the constant $c_n$ as
\cite{MatharArxiv0911,RaoJMO47}
\begin{equation}
c_n=-\frac{\Gamma[(3+\gamma)/2]}{2\pi^{3/2+\gamma}\Gamma(-\gamma/2)},
\label{eq.cn}
\end{equation}
in particular $c_n\approx 0.009693$ if $\gamma=2/3$.

\section{Line-of-Sight Path Integrals} \label{sec.2}

\subsection{Phase Structure Function}

We review the integration of isotropic structure functions ${\cal D}_n$ along the
line of sight
\cite[p. 293]{RoddierProgOpt19}\cite[(C1)]{FriedJOSA55}\cite[(7a)]{HufnagelJOSA54}.
The distance to the reference point
of the structure function is split into a horizontal vector
component ${\bf b}$ (baseline) and a 
scalar component $h$.
The phase $\varphi$ of the electromagnetic wave accumulated after transmission through
the turbulent layer is the product of the optical path length by the vacuum wave number $\bar k$,
and the optical path length is the path integral over the product of geometric
path length and refractive index $n$ along the atmospheric height $h$:
\begin{equation}
\varphi({\bf b}) =\bar k \int_0^{K/\sin a} n({\bf b},h)dh
\end{equation}
The scale height $K$ is the vertical thickness of the turbulent layer.
The air mass is $1/\sin a$ as a function of the star elevation angle $a$ above
the horizon.
The structure function of the phase is defined as
\begin{eqnarray}
{\cal D}_\varphi 
&=&\langle|\varphi(\mathbf 0)-\varphi(\mathbf b)|^2\rangle
.
\end{eqnarray}
The expectation value of the square is expanded with the binomial
theorem,
\begin{eqnarray}
{\cal D}_\varphi
&=&
\langle \left|\int_0^{K/\sin(a)}\bar kn(0,h)dh-\int_0^{K/\sin(a)}\bar k n(b,h)dh
\right|^2\rangle
\\
&=&
2\bar k^2\langle \left[
\int_0^{K/\sin(a)}n(0,h)dh
\int_0^{K/\sin(a)}n(0,h')dh'
-\int_0^{K/\sin(a)} n(0,h)dh
\int_0^{K/\sin(a)} n(b,h')dh'
\right]\rangle
\end{eqnarray}
We switch to relative and mean coordinates,
$\Delta h\equiv h-h'$ and $\bar h=(h+h')/2$,
with
inverse mapping $h=\bar h+\Delta h/2$, $h'=\bar h-\Delta h/2$,
\begin{eqnarray}
{\cal D}_\varphi
&=&
2\bar k^2
\Big[
\langle
\int_0^{K/(2\sin(a))}d\bar h
\int_{-2\bar h}^{2\bar h} d\Delta h
n(0,\bar h+\Delta h/2)
n(0,\bar h-\Delta h/2)
\rangle
\nonumber
\\
&& \quad
+
\langle
\int_{K/(2\sin(a))}^{K/\sin(a)} d\bar h
\int_{2[\bar h-K/\sin(a)]}^{2[K/\sin(a)-\bar h]} d\Delta h
n(0,\bar h+\Delta h/2)
n(0,\bar h-\Delta h/2)
\rangle
\nonumber
\\
&& \quad
-
\langle
\int_0^{K/(2\sin(a))}d\bar h
\int_{-2\bar h}^{2\bar h} d\Delta h
n(0,\bar h+\Delta h/2)
n(b,\bar h-\Delta h/2)
\rangle
\nonumber
\\
&& \quad
-
\langle
\int_{K/(2\sin(a))}^{K/\sin(a)} d\bar h
\int_{2[\bar h-K/\sin(a)]}^{2[K/\sin(a)-\bar h]} d\Delta h
n(0,\bar h+\Delta h/2)
n(b,\bar h-\Delta h/2)
\rangle
\Big].
\end{eqnarray}
In the first two terms of the right hand side, the
distance
between the locations in the arguments of the refractive indices
is $\Delta h$, because these are located on the same ray.
In the remaining two terms of the right hand side, the distance is between points
located on two different rays, horizontally translated by an amount $b$,
rays slanted by the altitude angle $a$ towards the horizontal. 
The location $(0,\bar h+\Delta h/2)$ on the first ray translates into Cartesian coordinates
$(h+\Delta h/2)(-\cos A\cos a,\sin A\cos a,\sin a)$ in a suitable
topocentric system with star azimuth $A$. The
location
$(b,\bar h-\Delta h/2)$ on the second ray has Cartesian coordinates
$(h-\Delta h/2)(-\cos A\cos a,\sin A\cos a,\sin a)
+b(-\cos A_b,\sin A_b,0)$
where $A_b$ is the baseline azimuth
\cite{MatharSAJ177}.
The square of the distance between these two locations is
$b^2+(\Delta h)^2-2b\Delta h \cos \theta$,
where $\cos\theta \equiv \cos a\cos(A_b-A)$ measures the angle $\theta$
between baseline and pointing.

\begin{figure}[hbt]
\includegraphics[scale=0.5,clip=true]{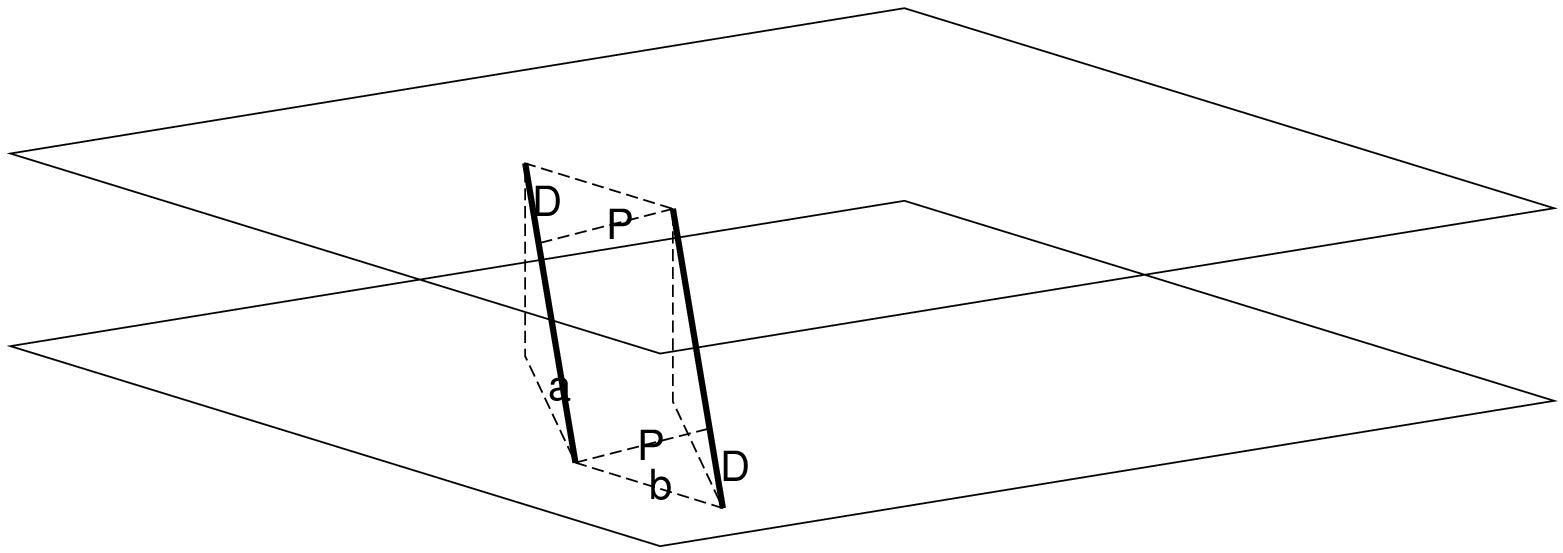}
\caption{
Sketch of geometric parameters. The sections of the two parallel rays
are the bold lines.
$a$ is the angle
between the rays and the horizontal.
The vertical distance between the top and bottom of the
layer is $K$; so the geometric length of each
ray is $K/\sin a$.
\label{fig.geo}
}
\end{figure}

In Figure
\ref{fig.geo},
$\theta$ is the angle between $b$ and what is called the delay $D$ in
astronomical interferometry,
\begin{equation}
D = b\cos\theta,
\end{equation}
and the third leg in this rectilinear triangle is the projected baseline
\begin{equation}
P=b\sin\theta.
\end{equation}

Reverse use of
\begin{equation}
{\cal D}_n=2 \langle n(0,h)^2\rangle-2\langle n(0,h)n(b,h)\rangle
\end{equation}
and insertion of the aforementioned distance
$b^2+(\Delta h)^2-2b\Delta h \cos \theta = P^2+(D-\Delta h)^2$
yields
\begin{eqnarray}
{\cal D}_\varphi
&=&
2\bar k^2
\langle
\int_0^{K/(2\sin(a))}d\bar h
\int_{-2\bar h}^{2\bar h} d\Delta h
n^2(0,\bar h)
\rangle
-\bar k^2
\int_0^{K/(2\sin(a))}d\bar h
\int_{-2\bar h}^{2\bar h} d\Delta h
{\cal D}_n(\Delta h)
\nonumber
\\
&&
+
2\bar k^2
\langle
\int_{K/(2\sin(a))}^{K/\sin(a)} d\bar h
\int_{2[\bar h-K/\sin(a)]}^{2[K/\sin(a)-\bar h]} d\Delta h
n^2(0,\bar h)
\rangle
-\bar k^2
\int_{K/(2\sin(a))}^{K/\sin(a)} d\bar h
\int_{2[\bar h-K/\sin(a)]}^{2[K/\sin(a)-\bar h]} d\Delta h
{\cal D}_n(\Delta h)
\nonumber
\\
&&
-
2\bar k^2
\langle
\int_0^{K/(2\sin(a))}d\bar h
\int_{-2\bar h}^{2\bar h} d\Delta h
n^2(0,\bar h)
\rangle
+\bar k^2
\int_0^{K/(2\sin(a))}d\bar h
\int_{-2\bar h}^{2\bar h} d\Delta h
{\cal D}_n(\sqrt{ P^2+(D-\Delta h)^2})
\nonumber
\\
&&
-
2\bar k^2
\langle
\int_{K/(2\sin(a))}^{K/\sin(a)} d\bar h
\int_{2[\bar h-K/\sin(a)]}^{2[K/\sin(a)-\bar h]} d\Delta h
n^2(0,\bar h)
\rangle
+\bar k^2
\int_{K/(2\sin(a))}^{K/\sin(a)} d\bar h
\int_{2[\bar h-K/\sin(a)]}^{2[K/\sin(a)-\bar h]} d\Delta h
{\cal D}_n(\sqrt{ P^2+(D-\Delta h)^2})
.
\end{eqnarray}
The four expectation values of the squared mean values cancel,
\begin{eqnarray}
{\cal D}_\varphi
&=&
-\bar k^2
\int_0^{K/(2\sin(a))}d\bar h
\int_{-2\bar h}^{2\bar h} d\Delta h
{\cal D}_n(\Delta h)
\nonumber
\\
&& \quad
-\bar k^2
\int_{K/(2\sin(a))}^{K/\sin(a)} d\bar h
\int_{2[\bar h-K/\sin(a)]}^{2[K/\sin(a)-\bar h]} d\Delta h
{\cal D}_n(\Delta h)
\nonumber
\\
&& \quad
+\bar k^2
\int_0^{K/(2\sin(a))}d\bar h
\int_{-2\bar h}^{2\bar h} d\Delta h
{\cal D}_n(\sqrt{ P^2+(D-\Delta h)^2})
\nonumber
\\
&& \quad
+\bar k^2
\int_{K/(2\sin(a))}^{K/\sin(a)} d\bar h
\int_{2[\bar h-K/\sin a]}^{2[K/\sin(a)-\bar h]} d\Delta h
{\cal D}_n(\sqrt{ P^2+(D-\Delta h)^2})
.
\end{eqnarray}
In the second and fourth term substitute $\bar h$ for
$\zeta = K/\sin(a) -\bar h$ to observe that their contributions double the values
of the first and third,
\begin{eqnarray}
{\cal D}_\varphi
&=&
-2\bar k^2
\int_0^{K/(2\sin(a))}d\bar h
\int_{-2\bar h}^{2\bar h} d\Delta h
{\cal D}_n(\Delta h)
\nonumber
\\
&& \quad
+2\bar k^2
\int_0^{K/(2\sin(a))}d\bar h
\int_{-2\bar h}^{2\bar h} d\Delta h
{\cal D}_n(\sqrt{ P^2+(D-\Delta h)^2})
\nonumber
\\
&=&
2\bar k^2
\int_0^{K/(2\sin(a))}d\bar h
\int_{-2\bar h}^{2\bar h} d\Delta h
\left[ {\cal D}_n(\sqrt{ P^2+(D-\Delta h)^2})
-
{\cal D}_n(\Delta h)
\right]
.
\label{eq.DphiDn}
\end{eqnarray}
Insertion of the Wiener spectrum (\ref{eq.D1d})
and substitutions
$\Delta h=Py$ and $\bar h=Pz/2$
of the integration variables transform this into
\begin{eqnarray}
{\cal D}_\varphi
&=&
16\pi c_nC_n^2\bar k^2 P
\int_0^\infty  \frac{f^2 df}{(f^2+f_0^2)^{(\gamma+3)/2}}
\int_0^{K/(2\sin a)}\!\!\!\!d\bar h
\int_{-2\bar h/P}^{2\bar h/P} dy
\left[
j_0(2\pi fPy)-j_0(2\pi fP\sqrt{1+(b\cos\theta/P -y)^2})
\right]
\label{eq.tmp}
\\
&=&
8\pi c_nC_n^2\bar k^2 P^2
\int_0^\infty  \frac{f^2 df}{(f^2+f_0^2)^{(\gamma+3)/2}}
\int_0^{K/(P\sin a)}dz
\int_{-z}^z dy
\left[
j_0(2\pi fPy)-j_0(2\pi fP\sqrt{1+(D/P -y)^2})
\right]
.
\label{eq.zyfInt}
\end{eqnarray}

\subsection{Review of the Standard Theory}

The standard proceeding assumes that
the contributions of the two spherical Bessel functions in (\ref{eq.tmp})
cancel well for large $y$. In consequence, the limits of the innermost integral are
extended to $\bar h/P\to \infty$, emitting a Bessel Function of order zero
\cite{FengAO41,LuckeAO46,VoitsekhoJOSAA12a}:
\begin{eqnarray}
\int_{-\infty}^{\infty} dy
j_0(2\pi fPy)
-
\int_{-\infty}^{\infty} dy
j_0(2\pi fP\sqrt{1+[b\cos\theta/P -y]^2})
\nonumber
\\
=
\frac{1}{2 fP}
-
\frac{1}{2 fP}J_0(2\pi fP)
.
\end{eqnarray}
This reduces the integration over $\bar h$ to a simple factor
\cite{Goodman,ChesnokovOC141}
\begin{eqnarray}
{\cal D}_\varphi
&=&
4\pi c_nC_n^2\bar k^2 \frac{K}{\sin a}
\int_0^\infty  \frac{f df}{(f^2+f_0^2)^{(\gamma+3)/2}}
[1-J_0(2\pi fP)]
.
\end{eqnarray}
The remaining frequency integrals are known in the literature
\cite[6.565.4]{GR}
the frequency-to-real-space matching factor (\ref{eq.cn})
is also inserted,
and the structure function for the von K\'arm\'an spectrum is condensed into the well-known
\cite{ConanJOSAA25,BeghiJOSAA25}
\begin{equation}
{\cal D}_\varphi(P)
=
- \bar k^2 \frac{K}{\sin a} C_n^2 P^{1+\gamma}
\frac{\sqrt{\pi}}{\Gamma(-\gamma/2)(\pi Pf_0)^{1+\gamma}}
\left[\Gamma\left(\frac{1+\gamma}{2}\right)
- 2(\pi Pf_0)^{(1+\gamma)/2}K_{(1+\gamma)/2}(2\pi Pf_0)\right]
,
\label{eq.vKar}
\end{equation}
where $K_.(.)$ are modified Bessel Functions.
\cite[9.6]{AS}
The factors that depend on the projected baseline (phase)
measured in units of the outer scale,
\begin{equation}
P_0 \equiv 2\pi f_0 P
,
\end{equation}
are the topic further below.
The Kolmogorov limit of infinite outer scale is
\begin{equation}
\lim_{f_0\to 0}
{\cal D}_\varphi(P)
=
\bar k^2 \frac{K}{\sin a} C_n^2 P^{1+\gamma} g(\gamma)
\label{eq.dKlim}
\end{equation}
where
\begin{equation}
g(\gamma)\equiv \sqrt{\pi} \frac{\Gamma(-1/2-\gamma/2)}{\Gamma(-\gamma/2)}
\approx 2.9143808
\label{eq.gofgamm}
\end{equation}
at $\gamma=2/3$. In overview, observations away from the zenith multiply the sale height
with the air mass, and employ an effective distance between the
two beams (which scales $\propto 5/3$) equal to the projected baseline.

\section{Correction for Finite Layer Thickness} \label{sec.3}

\subsection{Aim}

Quantitative understanding of effects of the finite thickness of turbulent
layers of air is of interest
where (i) the thickness is smaller than the entire
atmosphere, cases such as
subdivisions in models for multi-conjugated adaptive optics or
jet-streams, or 
(ii) where the lateral beam separation is comparatively large,
and the scaling down to small separations
is be done with a single, consistent structure constant,
for example comparing interferometric observations with atmospheric site monitor data
\cite{MaireAA448}

The core
of the following is to 
substitute the approximate calculation of Section 2.2 by transforming (\ref{eq.zyfInt}) into (\ref{eq.g0}).
The nature of the approximation reviewed above is sampling the difference of the
two structure functions (\ref{eq.DphiDn}) beyond the precise limits of the integral;
as the structure function is an increasing function of its argument,
the net effect of the approximation is an exaggeration of ${\cal D}_\varphi$,
at least in the case $D=0$ where the argument of the square root
remains larger than $\Delta h$.

We first constrain the turbulence model to the limit of infinite outer scale,
because the mathematics reduces to simpler special functions then
(Section 
\ref{sec.kolg}).
The generic case is phrased in terms of integrals over MacDonald Functions
in the main part (Section
\ref{sec.vkfull}).

\subsection{In the Kolmogorov Limit} \label{sec.kolg}

Taking the limit $f_0\to 0$, (\ref{eq.Dnf0}) is inserted into (\ref{eq.DphiDn}), 
\begin{eqnarray}
{\cal D}_\varphi
&=&
2C_n^2\bar k^2
\int_0^{K/(2\sin a)}d\bar h
\int_{-2\bar h}^{2\bar h} d\Delta h
\left\{ \left[P^2+(D-\Delta h)^2\right]^{\gamma/2}
-
|\Delta h|^\gamma
\right\}
\nonumber
\\
&=&
C_n^2\bar k^2 P^{1+\gamma}
\frac{K}{\sin a}
\frac{P\sin a}{K}
\int_0^{K/(P\sin(a))}dz
	\int_{-z}^{z} dy
\left\{ \left[1+(D/P-y)^2\right]^{\gamma/2}
-
|y|^\gamma
\right\}
.
\label{eq.zyI}
\end{eqnarray}

Binomial expansion of the integrand in hypergeometric series of $y$
transforms ${\cal D}_\varphi$ into 
\begin{equation}
{\cal D}_\varphi = C_n^2\bar k^2 P^{1+\gamma}\frac{K}{\sin a}g(\gamma,P_K)
\label{eq.kolg}
\end{equation}
where the dimensionless
\begin{equation}
P_K \equiv \frac{P \sin a}{K}
\end{equation}
is the
beam separation
in units of the path length, and where
\begin{equation}
g(\gamma,P_K)
=
\frac{P_K}{2}
\left[
F^{(1)}-2F^{(0)}+F^{(-1)}\right]
-\frac{1}{(1+\gamma/2)(1+\gamma)P_K^{1+\gamma}}
\label{eq.gambar}
\end{equation}
with
\begin{equation}
F^{(j)}\equiv \left(\frac{D}{P}+\frac{j}{P_K}\right)^2
\,_3F_2\left(\begin{array}{c}-\gamma/2,1/2,1\\ 3/2,2\end{array}\mid -\left(\frac{D}{P}+\frac{j}{P_K}\right)^2\right)
\label{eq.f32}
\end{equation}
generalizes (\ref{eq.gofgamm}) in terms of three generalized hypergeometric
functions.
Aspects of the implementation are discussed in Appendix
\ref{app.g}.

Figure
\ref{fig.gofm}
illustrates the magnitude of the effect.
Plotting $g$ as a function of $P_K$ focuses on the relative deviation
of the full computation in comparison with the standard formula (\ref{eq.dKlim});
non-horizontal lines indicate a nonlinearity between
${\cal D}_\varphi$ and $K/\sin a$ in (\ref{eq.kolg}).
In general, the phase fluctuations are smaller
than the factor 2.9 predicted
in the approximate (\ref{eq.gofgamm}),
which 
had been derived in
the limit $K/P\to \infty$ of infinite
layer thickness.
At optical and infrared
wavelengths, the thickness $K$
multiplied by the air mass is limited to the order of 10 km representing the entire
atmosphere (troposphere),
and $P$ is of the order of $2$ m for a single telescope up to $100$ m for
an interferometer. Whence $P_K$ is in the range $2\times 10^{-4}$ to $0.01$, and Figure
\ref{fig.gofm}
demonstrates that this finite layer thickness places $g$ in the range 2.3 to 2.7\@.
Long baseline interferometry ``drops'' power of ${\cal D}_\varphi$ through this
geometric sampling effect.

An increase of $D_\varphi$ is induced if $D/P= \cot \theta$ turns nonzero,
i.e., if the pointing direction is not perpendicular to the baseline.
This tendency is qualitatively understood from Figure
\ref{fig.geo},
since the contribution
at $D=0$ integrates correlations between the two beams at shortest distances $P$,
and
nonzero $D$ essentially
requires parallel sampling of the refractive index structure function
correlated over an effective distance $b>P$.

\begin{figure}[hbt]
\includegraphics[scale=0.5]{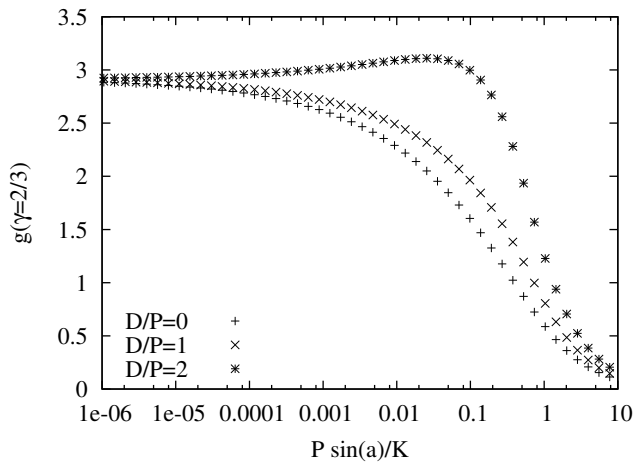}
\caption{
The factor (\ref{eq.gambar})
---results in the
limit of infinite outer scale---
as a function of $P_K$ (the ratio of the lateral distance over the path length)
for three examples of $D/P$.
\label{fig.gofm}
}
\end{figure}

\subsection{Von-K\'arm\'an Spectrum} \label{sec.vkfull}

The intent
is to evaluate (\ref{eq.zyfInt}),
\begin{eqnarray}
{\cal D}_\varphi
&=&
8\pi c_nC_n^2\bar k^2 P^{2+\gamma}
\int_0^\infty  \frac{f^2 df}{[f^2+(f_0 P)^2]^{(\gamma+3)/2}}
\int_0^{1/P_K}dz
\int_{-z}^z dy
\left[
j_0(2\pi fy)-j_0(2\pi f\sqrt{1+(D/P -y)^2})
\right]
\nonumber
\\
&=&
8\pi c_nC_n^2\bar k^2 P^{2+\gamma}
\int_0^\infty  \frac{f^2 df}{[f^2+(f_0 P)^2]^{(\gamma+3)/2}}
\int_0^{1/P_K}dz
\int_0^z dy
\left[
2j_0(2\pi fy)-j_0(2\pi f\sqrt{1+(D/P \mp y)^2})
\right]
\nonumber
\\
&=&
8\pi c_nC_n^2\bar k^2 P^{1+\gamma}\frac{K}{\sin a}
\int_0^\infty  \frac{f^2 df}{[f^2+(f_0 P)^2]^{(\gamma+3)/2}}
\int_0^{1/P_K}\!\!\!\!dy
[1-P_Ky]
\left[
2j_0(2\pi fy)-j_0(2\pi f\sqrt{1+[D/P \mp y]^2})
\right]
.
\label{eq.d3int}
\end{eqnarray}
with the convention that $\pm$ or $\mp$ indicate summation over
both signs in the corresponding term.
\subsubsection{First Ray}
The next step is to evaluate the previous equation.
It
contains a simpler term with a spherical Bessel function with argument $y$,
which will be called the contribution of the first ray, and another term
depending on $D/P$,
contribution of the second ray. Each splits into
two terms, one with and one without a factor $yP_K$.
Evaluating the $f$-integrals first, one term
is
\cite[3.771.5]{GR}\cite{vHaeringenMathComp39}
\begin{eqnarray}
&&
\int_0^\infty \frac{f^2df}{[f^2+(f_0 P)^2]^{(\gamma+3)/2}}
j_0(2\pi fy)
\nonumber
\\
&&
=
(2\pi y)^\gamma \int_0^\infty \frac{tdt}{[t^2+(2\pi yf_0 P)^2]^{(\gamma+3)/2}}\sin t
\nonumber
\\
&&
=
\frac{\sqrt{\pi}}{2^{1+\gamma/2}(f_0 P)^\gamma\Gamma(3/2+\gamma/2)}
(2\pi f_0P y)^{\gamma/2}K_{\gamma/2}(2\pi yf_0 P)
.
\label{eq.fint}
\end{eqnarray}
Its $y$-integral is
\begin{equation}
\int_0^{1/P_K}
(P_0y)^{\gamma/2}K_{\gamma/2}(P_0 y)dy
=
\frac{1}{P_0} V_\nu(P_0/P_K)
,
\label{eq.11st}
\end{equation}
at
\begin{equation}
\nu\equiv \gamma/2,
\end{equation}
where we define
\cite[6.561.4]{GR}
\begin{eqnarray}
V_\nu(x)\equiv \int_0^x t^\nu K_\nu(t)dt
= 2^{\nu-1}\sqrt{\pi}\Gamma(1/2+\nu)x
\left[K_\nu(x)\mathbf{L}_{\nu-1}(x)+K_{\nu-1}(x)\mathbf{L}_\nu(x)\right],
\label{eq.Vdef}
\end{eqnarray}
with $\mathbf{L}$ representing the modified Struve functions
\cite[12.2]{AS}.
The other contribution of the first ray is
\cite[11.3.27]{AS}
\begin{eqnarray}
&&
\int_0^{1/P_K} y (P_0 y)^{\gamma/2}K_{\gamma/2}(P_0 y) dy
=
-\frac{1}{P_0^2}
\left[
(P_0/P_K)^{1+\nu}K_{1+\nu}(P_0/P_K) -\Gamma(1+\nu)2^\nu \right]
.
\label{eq.12nd}
\end{eqnarray}

\subsubsection{Second Ray}
The companion of (\ref{eq.fint}) from the second ray is
\begin{eqnarray}
&&
\int_0^\infty \frac{f^2 df}{[f^2+(f_0P)^2]^{(\gamma+3)/2}}j_0(2\pi f\sqrt{1+[y\mp D/P]^2})
\nonumber\\
&&
=
\frac{\sqrt{\pi}}{2^{1+\gamma/2}(f_0 P)^\gamma\Gamma(3/2+\gamma/2)}
[P_0\sqrt{1+(y\mp D/P)^2}]^\nu K_\nu(P_0 \sqrt{1+[y\mp D/P]^2})
.
\end{eqnarray}
Its $y$-integral is
\begin{eqnarray}
\int_0^{K/(P \sin a)} (1-P_Ky)
[P_0\sqrt{1+(y\mp D/P)^2}]^\nu K_\nu(P_0 \sqrt{1+[y\mp D/P]^2})
dy
\nonumber\\
=
\int_{\mp D/P}^{1/P_K\mp D/P} (1-P_Ky\mp P_K\frac{D}{P})
[P_0\sqrt{1+y^2}]^\nu K_\nu(P_0 \sqrt{1+y^2})
dy
.
\end{eqnarray}
A substitution $P_0\sqrt{1+y^2}=t$ calls to split
the interval into the regions $y>0$ and $y<0$, which
is qualified with
the step-function,
$\Theta(x)\equiv 1$ if $x>0$ and $\Theta(x)\equiv 0$ if $x<0$.
The contribution from positive $y$ to the integral is
\begin{eqnarray}
\frac{\Theta(1/P_K\mp D/P)}{P_0}
\bigg[
(1\mp P_K\frac{D}{P})
U_\nu(t,P_0 )
+
\frac{P_K}{P_0}
t^{\nu+1}K_{\nu+1}(t)
\bigg]
_{t=P_0\sqrt{1+\max^2(0,\mp D/P)}}^{P_0\sqrt{1+(1/P_K\mp D/P)^2}}
,
\label{eq.2y1}
\end{eqnarray}
where
\begin{equation}
U_\nu(z,u)\equiv
\int_u^z \frac{t^{1+\nu}}{\sqrt{t^2-u^2}}K_\nu(t)dt
\label{eq.Udef}
\end{equation}
is an integral
described
in Appendix
\ref{app.U}.
The contribution from
negative $y$
is
\begin{equation}
\frac{\Theta(\pm D/P)}{P_0}
\bigg[
(1\mp P_K\frac{D}{P})
U_\nu(t,P_0 )
-
\frac{P_K}{P_0}
t^{\nu+1}K_{\nu+1}(t)
\bigg]
_{t=P_0\sqrt{1+\min^2(0,1/P_K\mp D/P)}}^{P_0\sqrt{1+(D/P)^2}}
.
\label{eq.2y2}
\end{equation}

\subsubsection{Combined Master Equation}

Collecting (\ref{eq.11st}), (\ref{eq.12nd}), (\ref{eq.2y1}) and (\ref{eq.2y2})
puts  (\ref{eq.d3int}) into the format
\begin{equation}
{\cal D}_\varphi = C_n^2\bar k^2P^{1+\gamma}\frac{K}{\sin a}g(\gamma,P_0,P_K),
\end{equation}
with
\begin{eqnarray}
g(\gamma, P_0, P_K)
&=&
-\frac{2^{1+\nu}}{\Gamma(-\nu)P_0^{1+\gamma}}
\bigg\{
2V_{\nu}(P_0/P_K)
\nonumber \\
&&
+
2\frac{P_K}{P_0}
\left[(\frac{P_0}{P_K})^{1+\nu}K_{1+\nu}(P_0/P_K)-\Gamma(1+\nu)2^\nu\right]
\nonumber \\
&&
-\Theta(1/P_K\mp D/P)
\bigg[
(1\mp P_K\frac{D}{P})
U_\nu(t,P_0 )
+
\frac{P_K}{P_0}
t^{\nu+1}K_{\nu+1}(t)
\bigg]
_{t=P_0\sqrt{1+\max^2(0,\mp D/P)}}^{P_0\sqrt{1+(1/P_K\mp D/P)^2}}
\nonumber\\
&&
-
\Theta(\pm D/P)
\bigg[
(1\mp P_K\frac{D}{P})
U_\nu(t,P_0 )
-
\frac{P_K}{P_0}
t^{\nu+1}K_{\nu+1}(t)
\bigg]
_{t=P_0\sqrt{1+\min^2(0,1/P_K\mp D/P)}}^{P_0\sqrt{1+(D/P)^2}}
\bigg\}
\label{eq.g0}
\end{eqnarray}
at $\nu=\gamma/2$. A summation over
the upper and lower
sign is still implied for the terms with the two $\Theta$-function factors.
[Note that there are massive cancellations between the
group of terms containing $V_\nu$ and $U_\nu$,
and also between the other terms depending on $K_{1+\nu}$,
which can be traced back to
the differences of the spherical Bessel functions $j_0$.]

This is the main result.
Equation (\ref{eq.vKar})
represents the concerted limit $P_K\to 0$, $D/P\to 0$,
equation (\ref{eq.gofgamm})
the limit $P_0\to 0$, $P_K\to 0$, $D/P\to 0$,
equation (\ref{eq.gambar})
the limit $P_0\to 0$, and
equation
(\ref{eq.gord0})
the limit $P_0\to 0$, $D/P\to 0$.

\begin{figure}[hbt]
\includegraphics[scale=0.5]{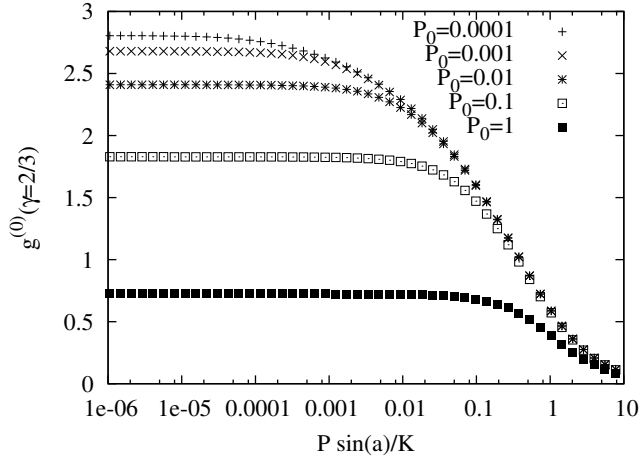}
\caption{
The factor (\ref{eq.g0}) as a function of the dimensionless $P_K$
for five different dimensionless measures $P_0$ of the outer scale, with $D/P=0$.
\label{fig.gofmVK}
}
\end{figure}

Figure
\ref{fig.gofmVK}
shows again $g^{(0)}$ to point out at which
lateral distances the full theory starts to deviate from the straight lines
predicted by the standard theory. The upper index indicates that
this is the contribution from the zeroth order of an expansion in powers of $(D/P)$, as in 
App.\ \ref{app.g}.

A numerical example:
The curve $P_0=0.1$ is equivalent to a separation $P=40$ cm if the outer scale
is $1/f_0=25$ m.
The standard formula (\ref{eq.vKar}) assumes a value of $g=1.83$ at the left
edge of the plot; the refined calculation reduces this by 25\%
to $g=1.37$ if $P\sin a/K=0.137$, i.e, if the path length through
the layer is
$K/\sin a\approx 2.9$ m.
As layers are thicker in practise, the more advanced calculation
has
little impact here. This seems to contradict the
importance claimed in Section
\ref{sec.kolg},
but the following mechanism conciliates both views:
As $P_0$ increases, i.e., as the outer scale shrinks,
the curves in Figure
\ref{fig.gofmVK}
remain horizontal over an increasingly wide initial range on the
$P_K$ axis.
The standard formula
remains accurate down to thin turbulent layers,
because the
clamping of ${\cal D}_n$ introduced by  the outer scale reduces the importance
of correlations over large (including large vertical) scales.
The additional
strict cut
within the refined calculation targets smaller
values of ${\cal D}_n$, so the relative effect becomes less important.

\begin{figure}[hbt]
\includegraphics[scale=0.5]{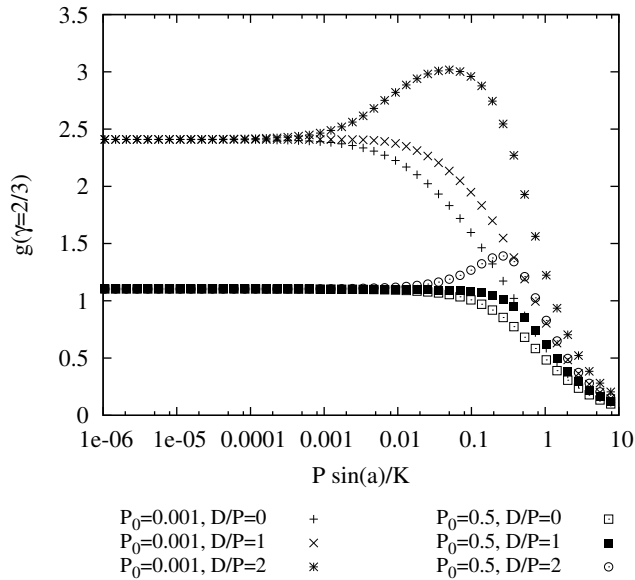}
\caption{
The factor (\ref{eq.g0}) as a function of the dimensionless $P_K$
for two different dimensionless measures $P_0$ of the outer scale and three
different
ratios $D/P$.
\label{fig.gofmVKDP}
}
\end{figure}

Inclusion of nonzero ``shear'' ratios $D/P$ leads to Figure
\ref{fig.gofmVKDP},
which shows no other features than those expected by combining
the shapes of Figures
\ref{fig.gofm}
and
\ref{fig.gofmVK}:
a hump appears in $g$ if the delay $D$ occupies a major fraction of the path length,
and $g$ falls off with an approximate $-5/6$th power at large $P_0$.

\section{Summary}

We developed a formula of the phase structure function depending on the spectral
index, on the lateral separation, on the air mass and
layer thickness, and on the outer scale, which rigorously
samples the von-K\'arm\'an statistics of isotropic homogeneous turbulence
along two parallel rays.

\bibliographystyle{SAJ}
\bibliography{all}

\appendix

\section{Kolmogorov Limit}\label{app.g}

The generalized Hypergeometric Function (\ref{eq.f32}) reduces effectively
to Gaussian Hypergeometric Functions
\cite{GottschalkJPA21,RainvilleBAMS51}:
\[
_3F_2\left(\begin{array}{c}1/2,1,-\gamma/2 \\ 3/2,2\end{array}\mid z\right)
=
2\,_2F_1\left(\begin{array}{c}-\gamma/2,1/2 \\ 3/2\end{array}\mid z\right)
-\,_2F_1\left(\begin{array}{c}-\gamma/2,1 \\ 2\end{array}\mid z\right)
.
\]
A useful Taylor expansion of (\ref{eq.gambar}) in powers of $D/P$ is
\begin{equation}
g(\gamma,P_K)= g^{(0)}+g^{(2)}+g^{(4)}+\cdots
\end{equation}
with constant order
\begin{equation}
g^{(0)}(\gamma,P_K)
=
-\frac{1}{1+\gamma/2}
\left[
P_K
\left\{\,_2F_1\left(\begin{array}{c}-1/2,-\gamma/2-1\\1/2\end{array}\mid -\frac{1}{P_K^2}\right)-1\right\}
+ \frac{1}{(1+\gamma)P_K^{1+\gamma}}\right]
,
\label{eq.gord0}
\end{equation}
quadratic order
\begin{equation}
g^{(2)} =
P_K (\psi^{\gamma/2}-1) (\frac{D}{P})^2
,
\end{equation}
and biquadratic order
\begin{equation}
g^{(4)} =
P_K
\gamma\frac{\psi^{\gamma/2}(2+\gamma\psi-\gamma-\psi)-\psi^2}{12\psi^2}
(\frac{D}{P})^4
,
\end{equation}
where
\begin{equation}
\psi \equiv 1+1/P_K^2.
\end{equation}

\section{Integral $U$}\label{app.U}

If the upper limit $z$ of the integral (\ref{eq.Udef}) is large,
it is numerically advantageous to use the complementary
\cite[6.592.12]{GR}
\begin{equation}
U_\nu(z,u)
= 
\sqrt{\pi/2}u^{\nu+1/2}K_{\nu+1/2}(u)
-\int_z^\infty 
\frac{t^{\nu+1}}{\sqrt{t^2-u^2}}K_\nu(t)dt
.
\end{equation}
Further binomial expansion in a power series of $u/z$,
\begin{equation}
\int_z^\infty \frac{t^{\nu+1}}{\sqrt{t^2-u^2}}K_\nu(t) dt
=
\sum_{l\ge 0}{-1/2 \choose l}(-u^2)^l\int_z^\infty t^{\nu-2l}K_\nu(t)dt
\end{equation}
leads to a set of integrals where the parameter $l$ is recursively reduced
by partial integration if $l>0$
\cite{MillerJMAA145},
\begin{equation}
\int_z^\infty t^{\nu-2l}K_\nu(t)dt
= \frac{z^{\nu-2l+1}}{2l-1}K_\nu(z)
+
\frac{1}{1-2l}\int_z^\infty t^{-2(l-1)}t^{\nu-1} K_{\nu-1}(t)dt
\end{equation}
until the case $l=0$ is
covered by (\ref{eq.Vdef}) via
\cite[6.561.16]{GR}
\begin{equation}
\int_z^\infty t^{\nu}K_\nu(t)dt = \sqrt{\pi}2^{\nu-1}\Gamma(\nu+1/2)-V_\nu(z).
\end{equation}
On the other hand, if $z$ is not larger than approximately $1.8u$, the repeated partial integration
represents a converging series
\begin{equation}
U_{\nu}(z,u)
=
\sum_{k\ge 0}
\frac{1}{1\cdot 3\cdot 5\cdots (2k-1)(2k+1)}(z^2-u^2)^{k+1/2} z^{\nu-k}K_{\nu-k}(z)
.
\end{equation}

\end{document}